\documentstyle[twoside,11pt,draft]{article}
\newcommand{\be}{\begin{equation}}
\newcommand{\ee}{\end{equation}}
\newcommand{\bra}[1]{\mbox{$\langle\, #1 \mid$}}
\newcommand{\ket}[1]{\mbox{$\mid #1\,\rangle$}}
\newcommand{\pro}[2]{\mbox{$\langle\, #1 \mid #2\,\rangle$}}
\newcommand{\expec}[1]{\mbox{$\langle\, #1\,\rangle$}}
\newcommand{\expecl}[1]{\mbox{$\left\langle\, 
            \strut\displaystyle{#1}\,\right\rangle$}}

\title{Semiclassical Collapse of a Sphere of Dust}
\author{Roberto Casadio\thanks{e--mail: Casadio@bologna.infn.it} 
\ and Giovanni Venturi\thanks{e--mail: Armitage@bologna.infn.it}\\
 \\
{\em Dipartimento di Fisica, Universit\`a di
Bologna} \\
{\em and} \\
{\em Istituto Nazionale di Fisica Nucleare, 
Sezione di Bologna, Italy}}
\begin{document}
\baselineskip 4.0ex
\begin{titlepage}
\pagestyle{empty}
\maketitle
\begin{abstract}
The semiclassical collapse of a homogeneous sphere of dust
is studied.
After identifying the independent dynamical variables, 
the system is canonically quantised and coupled equations
describing matter (dust) and gravitation are obtained.
The conditions for the validity of the adiabatic 
(Born--Oppenheimer) and semiclassical approximations
are derived.
Further on neglecting back--reaction effects, it is shown
that in the vicinity of the horizon and inside the dust
the Wightman function for a conformal scalar field
coupled to a monopole emitter is thermal at the
characteristic Hawking temperature.
\end{abstract}
\end{titlepage}
%
\pagestyle{plain}
\raggedbottom
\setcounter{page}{1}
\section{Introduction}
The canonical quantisation of general relativistic isotropic 
systems carried out in suitably chosen variables leads
to the whole dynamics being determined by the Hamiltonian
constraint \cite{dewitt} of the Arnowitt, Deser and Misner 
(ADM) construction \cite{adm}.
Such an approach is particularly useful if one wishes to study
the semiclassical regime of a system of self gravitating 
matter \cite{brout}.
In this note we apply the ADM formalism to
a model describing the collapse of a spherical object 
leading to a black hole. 
Further since one of the more intriguing 
aspects of quantum field theory in the curved background 
generated by a black hole is thermal Hawking radiation
\cite{hawking}, we examine how it arises within such an
approach.
\par
In section~\ref{sec_1} we briefly review the classical dynamics of an
isotropic perfect fluid with zero pressure (dust) represented
by a scalar field $\phi$ \cite{madsen,lund}.
In particular we observe that with a suitable gauge choice for
the radial coordinate \cite{lund}, the supermomentum constraint 
can be solved at the classical level and one 
is left with a Hamiltonian whose flow lines are the 
well known inhomogeneous Tolman solutions \cite{tolman}.
These reduce to the more specific Oppenheimer--Snyder 
model \cite{oppenheimer} if one also assumes 
homogeneity (subsection~\ref{homo}).
\par
In contrast with a previous treatment \cite{lund} we do not 
impose the temporal gauge $\phi=-t$ which amounts to using the 
(classical) dust degree of freedom as a clock for the system and 
also restricts the solutions to the homogeneous case only.
In fact we shall introduce time only after having 
quantised the remaining Hamiltonian constraint
obtaining the Wheeler--DeWitt equation.
This is not a trivial task and in subsection~\ref{homo}
we mention briefly the main conceptual problems one encounters:
the operator ordering, the singularity at the origin and
the absence of time.
\par
In order to overcome all of these difficulties,
in section~\ref{semicl} we illustrate a decomposition
scheme for the Wheeler--DeWitt equation of the 
Oppenheimer--Snyder model in analogy with composite
systems such as molecules which have two time (or mass)
scales. 
The total wave function is factorised into a matter
and a gravitational part \cite{brout} and on neglecting
fluctuations (Born--Oppenheimer or adiabatic approximation),
matter will follow gravity in the semiclassical limit for 
the latter.
\par
Finally, in section~\ref{radiation} we introduce a massless 
scalar field conformally coupled to gravity and we allow it 
to interact locally with a monopole emitter localised
near the horizon in coincidence with the last shell of the
collapsing dust.
This is done with the aim of obtaining information about
the emitted radiation (temperature and intensity) as the
last shell approaches the horizon.
\par
We use units for which $c=k_{Boltzmann}=1$, $\kappa\equiv 8\,\pi\,G$,
the Planck length is then $\ell_p\equiv\sqrt{\kappa\,\hbar}$
and the Planck mass is $m_p=\hbar/\ell_p$.
\setcounter{equation}{0}
\section{The isotropic sphere of dust}
\label{sec_1}
The most general three dimensional isotropic metric can be written
as:
\begin{equation}
^{(3)}ds^2=e^{2\,\nu}\,d\zeta^2+e^{2\,\lambda}\,d\Omega^2
\ ,
\label{3g_in1}
\end{equation}
where $d\Omega^2$ is the surface element of a unit 2-sphere,
$\nu=\nu(\zeta,\sigma)$ and $\lambda=\lambda(\zeta,\sigma)$ are the
gravitational degrees of freedom which depend only on the
internal time variable $\sigma$ and the radial coordinate $\zeta$.
A real massive scalar field $\phi=\phi(\zeta,\sigma)$ can be related 
to the dust 4-velocity $U^\alpha$ through
$U_\alpha=\phi_{,\alpha}$
and Hamilton's equations of motion
for the self--gravitating sphere of dust can be obtained by
variation of the action functional:
\begin{equation}
S=4\,\pi\,\int d\sigma\,d\zeta\,\left[
\pi_\nu\,{\partial\nu\over\partial\sigma}
+\pi_\lambda\,{\partial\lambda\over\partial\sigma}
+\pi_\phi\,{\partial\phi\over\partial\sigma}
-N\,{\cal H}-N_1\,{\cal H}^1\right]
\ ,
\end{equation}
with respect to $\nu$, $\lambda$, $\phi$, their
conjugate momenta $\pi_\nu$, $\pi_\lambda$ and
$\pi_\phi=-\sqrt{|g|}\,\mu\,U^0$.
Here $\mu=\mu(\zeta,\sigma)$ is the rest mass density of the dust,
$g$ is the determinant of the 4-metric whose pull--back 
on the 3-surface $\Sigma_\sigma$ of constant $\sigma$ is given in 
Eq.~(\ref{3g_in1}), $N$ and $N^1$ are the lapse and radial shift 
functions.
\par
In the ADM formalism, $N$ and $N_1$ play the role of Lagrange multipliers
for the Hamiltonian density ${\cal H}$
and the radial supermomentum ${\cal H}^1$ 
respectively,
leading to the Hamiltonian and radial supermomentum constraints
\cite{lund,bcmn}:
\begin{eqnarray}
{\cal H}&=&e^{-\nu-2\,\lambda}\,
\left[{\kappa\over 8}\,\pi_\nu^2-{\kappa\over 4}\,\pi_\nu\,\pi_\lambda
+{2\over\kappa}\,e^{4\,\lambda}\,\left(
2\,\lambda''-2\,\lambda'\,\nu'+3\,(\lambda')^2-e^{2\,(\nu-\lambda)}
\right)\right] \nonumber \\
& &+\pi_\phi\,\left[1+e^{-2\,\nu}\,(\phi')^2\right]^{1/2}=0
\label{h=0} \\
& & \nonumber\\
{\cal H}^1&=& -e^{-2\,\nu}\,\left(
{\pi_\nu}'-\nu'\,\pi_\nu-\lambda'\,\pi_\lambda\right)
-e^{-2\,\nu}\,\pi_\phi\,\phi'=0
\label{chi=0}
\ ,
\end{eqnarray}
where a prime denotes the radial derivative
$\partial/\partial \zeta$ and the above 
are secondary constraints in Dirac's language and arise
because of the freedom of scaling the time variable $\sigma$ and the
radial coordinate $\zeta$ on $\Sigma_\sigma$.
\par
The residual gauge freedoms allow one to choose both
the {\em comoving gauge\/} $U^\alpha=\delta^\alpha_{\ 0}$, that is 
$\phi'=0$ and then the scalar field is a function of time only,
and the {\em radial gauge} $\lambda'\,e^{\lambda-\nu}=\zeta$
which is possible if $\lambda'\,e^{\lambda-\nu}$,
considered as a new canonical variable, is a monotonic
positive regular function of $\zeta$ at each value of $\sigma$.
One then has $\lambda$ as the only gravitational degree 
of freedom and on solving the 
constraints Eqs.~(\ref{h=0}) and (\ref{chi=0})
one obtains a final Hamiltonian which does not contain any spatial
derivatives so that the dynamics for different radii (shells)
decouples.
This is a consequence of the fact that dust is not subject
to other interactions except gravity and Birkhoff's theorem
ensures that each shell at constant $\zeta$ probes
only the total mass in its interior \cite{mtw}.
\par
After performing a further canonical transformation
$r=e^\lambda$ from the Hamiltonian equations
of motion one obtains a single equation:
\begin{equation}
\dot r^2+2\,r\,\ddot r+(1-\zeta^2)=0
\ .
\label{ham}
\end{equation}
Here and in the following a dot will denote a derivative 
with respect to proper time $d\tau\equiv N\,d\sigma$.
It is also useful to define $\rho\equiv\sqrt{(1-\zeta^2)/\epsilon}$,
with:
\be
\epsilon=\left\{\begin{array}{ll}
+1 & \ \ \ \   0\le \zeta<1\\
0 &  \ \ \ \  \zeta=1 \\
-1&  \ \ \ \   \zeta>1
\ .
\end{array}\right.
\ee
In terms of this new radial coordinate the interior of 
the sphere of dust is given by 
$0\le\rho\le\rho_{_0}$, where $\rho_{_0}$ is the
coordinate of the outer surface of the sphere:
$0<\rho_{_0}<1$ for $\epsilon=+1$, 
$\rho_{_0}>0$ for $\epsilon=0,-1$.
Of course, for $\epsilon=0$ the above definition is
undefined (it corresponds to flat space) 
and we simply have $\zeta(\rho)=1$ for 
$0\le\rho\le\rho_{_0}$.
\subsection{The homogeneous case and matching with external
Schwarzschild metric}
\label{homo}
The solutions to Eq.~(\ref{ham}) were obtained
long ago \cite{tolman} and, 
since the density $\kappa\,\mu(\rho,\tau)=F'/(r^2\,r')$
is determined only by the arbitrary function $F(\rho)$,
we can restrict our analysis to the case in which the dust is
homogeneous at each instant of proper time $\tau$ 
by taking $F(\rho)=2\,\rho^3\,K_0$,
where $K_0$ is a constant.
The equations of geodesic motion then become:
\begin{eqnarray}
r&=&\rho\,K_0\,\partial_\eta h_\epsilon(\eta)\equiv
\rho\,K_{cl}(\eta)
\nonumber \\
& & \label{hom}\\
\tau&=&\tau_{_0}\pm K_0\,h_\epsilon(\eta)
\ ,
\nonumber
\end{eqnarray}
where $\tau_{_0}$ is another constant and:
\be
h_\epsilon(\eta)=\left\{\begin{array}{ll}
\eta-\sin\eta &\ \ \ \ \ \epsilon=+1 \\
& \\
\eta^3/6      &\ \ \ \ \ \epsilon=0  \\
& \\
\sinh\eta-\eta&\ \ \ \ \ \epsilon=-1
\ ,
\end{array}\right.
\label{h}
\ee
with $\eta$ a new time variable related to the proper time through
$d\eta={\rho\over r}\,d\tau$.
The density is now given by:
\be
\kappa\,\mu(\eta)={6\,K_0\over K_{cl}^3(\eta)}
\ ,
\label{densita}
\ee
and the line element by:
\be
ds^2=K_{cl}^2(\eta)\left[-d\eta^2+{d\rho^2\over 1-\epsilon\,\rho^2}
+\rho^2\,d\Omega^2\right]
\ .
\label{g_hom}
\ee
One then sees that the parameter $\eta$ is the
{\em conformal time\/} and the portion of space--time filled
with dust has the structure of a homogeneous Robertson--Walker 
manifold \cite{mtw}.
The case $\epsilon=+1$ corresponds to a sphere starting from
its maximum radius $r=2\,\rho\,K_0$ at $\eta=\pi$ which
finally collapses to a point at $\eta=0$ while
the other two cases ($\epsilon=0,-1$) start from an initial
radius at $\eta=\eta_{_0}>0$ and again collapse at $\eta=0$.
\par
We may now consider the exterior metric:
the simplest assumption is to have vacuum
outside the sphere of dust, so that isotropy and Birkhoff's theorem
lead to the unique Schwarzschild solution \cite{mtw}
which must be matched at the surface $r_{_0}$ of the dust sphere.
This requires that each shell of dust move along a geodesic in
both metrics \cite{mtw,israel} implying that a shell at
$\rho\le\rho_{_0}$ moves along $r=\rho\,K_{cl}(\eta)\le r_{_0}$
in a local metric of the Schwarzschild type  
with a mass parameter $M_{_G}(\rho)=\,K_0\,\rho^3/\kappa$.
The conformal time $\eta_{_H}(\rho)$ at which 
each shell crosses its own Schwarz\-schild radius is then given
by $r_{_H}(\rho)\equiv 2\,M(\rho)=\rho\,K_{cl}(\eta_{_H})$,
so that from the point of view of a distant observer
the first horizon to form is the outer one 
(and therefore the only relevant one)
located at:
\be
r_{_H}\equiv 2\,M=2\,K_0\,{\rho_{_0}}^3
\ ,
\label{M}
\ee
which is the Schwarzschild radius of the sphere.
This whole picture is also known as the {\em Oppenheimer--Snyder
model\/} \cite{oppenheimer}.
\par
One may further express the interior conformal time
$\eta$ as a function of the external Schwarzschild time $t$ for the
geodesic trajectory at the surface $r=r_{_0}$.
This is rather complicated \cite{mtw}, 
but we are only interested in the asymptotic relations for 
$r_{_0}\to r_{_H}$.
In such a limit one has: 
\be
t\simeq\left\{\begin{array}{ll}
-2\,M\,\ln\{\tan(\eta/2)-\tan(\eta_{_H}/2)\}
&\ \ \ \epsilon=+1 \\
& \\
-2\,M\,\ln(\eta-\eta_{_H})
&\ \ \ \epsilon=0 \\
& \\
-2\,M\,\ln\{\tanh(\eta/2)-\tanh(\eta_{_H}/2)\}
&\ \ \ \epsilon=-1 
\ .
\end{array}\right.
\label{t(eta)}
\ee
Eqs. (\ref{t(eta)}) are of course singular at $\eta=\eta_{_H}$
since we know that $t$ must diverge when the outer shell approaches
the horizon, whereas nothing happens to a comoving
observer.
For all the above three cases the equations can be inverted
to give:
\be
\eta-\eta_{_H}\simeq C_\epsilon\,e^{-t/2\,M}
\simeq C_\epsilon\,\left(1-{2\,M\over r_{_0}(t)}\right)
\ ,
\label{eta(t)}
\ee
with $C_\epsilon$ a positive constant.
The last factor is the square of the redshift for a photon
escaping from near the black hole.
\par
We may summarize our results as follows:
the Einstein--Hilbert action for the metric (\ref{g_hom}) is given by:
\be
S_{_G}=-{V_s\over2\,\kappa}\,\int d\tau\left[
K\,\dot K^2-\epsilon\,K\right]
\ ,
\label{s_g}
\ee
where  
$V_s=4\,\pi\,\int_0^{\rho_{_0}} d\rho\,\rho^2/\sqrt{1-\epsilon\,\rho^2}$
is the spatial volume of the sphere and, in order to reproduce the 
solutions Eq.~(\ref{hom}), the action for the dust must be:
\be
S_{_M}=-{V_s\over\kappa}\,K_0\,\int d\tau
\ ,
\label{s_m}
\ee
with a final Hamiltonian constraint:
\be
-{1\over2}\,\left(\kappa\,
{\pi_{_K}^2\over K}+{\epsilon\over\kappa}\,K
\right)+{K_0\over\kappa}=0
\ ,
\label{h=0_frw}
\ee
\par
Let us note that Eq.~(\ref{h=0_frw}) has been
obtained {\em after} having imposed the condition that matter
satisfy the (classical) equation of state for dust (that is 
a perfect fluid with zero pressure) and {\em after\/} having
solved for the classical radial momentum 
constraint Eq.~(\ref{chi=0}).
This in electrodynamics would correspond to a choice of gauge
in order to eliminate unphysical degrees of freedom.
On having identified the independent degrees of freedom
one may proceed to quantise canonically Eq.~(\ref{h=0_frw}) 
by replacing
$\pi_{_K}\mapsto\hat\pi_{_K}=-i\,\hbar\,{\partial\over\partial K}$.  
However one must also face the following issues:
\begin{enumerate}
\item
{\em Operator ordering}.
The formal mapping does not determine 
the quantum kinetic term uniquely: 
\be
{\pi_{_K}^2\over K}\mapsto 
{1\over\hat K^a}\,\hat\pi_{_K}^m\,{1\over\hat K^b}\,\hat\pi_{_K}^n\,
{1\over\hat K^c}
\ ,
\ee 
where $a$, $b$, $c$, $m$ and $n$ can be arbitrary integers such that 
$a+b+c=1$ and $m+n=2$ with $m$ and $n$ positive,
since in all cases the same classical limit is obtained.
\item
{\em Singularity at the origin}.
The wave function $\Psi=\Psi(K,\phi)$ satisfying the quantised 
Wheeler--DeWitt equation must be zero at the origin ($K=0$).
\item
{\em Definition of time}.
Eq.~(\ref{h=0_frw}) does not contain a time variable,
so one must face the problem of the introduction of
time in order to recover classical motion.
\end{enumerate}
A consistent scheme solution to the above difficulties
which also allows for the quantisation of matter
is discussed in section~\ref{semicl}.
\setcounter{equation}{0}
\section{The semiclassical collapse}
\label{semicl}
In the previous section we briefly reviewed for completeness the 
classical description of the evolution of a sphere of dust.
We finally limited ourselves to the homogeneous case 
which we shall further analyse
(a generalization to inhomogeneous collapse can be done by
considering separately shells at different radii \cite{lund}).
\par
As a guide for quantisation one has that in the semiclassical limit
one must recover classical motion from some quantum state of 
gravity and matter.
For our purpose it is convenient to consider a system consisting of
a portion of a Robertson--Walker space--time plus a
scalar field $\phi=\phi(\tau)$ whose total classical action is 
given by:
\be
{S\over V_s}={1\over2}\,\int d\tau\,\left[
-{1\over\kappa}\,\left(K\,{\dot K}^2-\epsilon\,K\right)
+K^3\,\left({\dot\phi}^2-{\mu}^2\,{\phi}^2\right)
\right]
\ ,
\label{s1}
\ee
where $\mu=m/\hbar$ is the inverse of the Compton 
wavelength of the field $\phi$.
On using the equations of motion for dust \cite{madsen}
and identifying the density as in Eq.~(\ref{densita})
the above action becomes:
\be
{S\over V_s}=-{1\over2\,\kappa}\,\int d\tau\,\left[
K_{cl}\,{\dot K_{cl}}^2-\left(\epsilon\,K_{cl}-2\,K_{_0}\right)
\right]\equiv{S_{cl}\over V_s}
\ ,
\label{s_cl}
\ee
where $K_{cl}$ is defined in Eq.~(\ref{hom}).
This is in agreement with what has been obtained in 
section~\ref{sec_1}:
$S_{cl}=S_{_G}+S_{_M}$ where $S_{_G}$ is given by Eq.~(\ref{s_g})
and $S_{_M}$ by Eq.~(\ref{s_m}).
\par
The Hamiltonian obtained from the action in Eq.~(\ref{s1}) 
is:
\be
H=-{1\over2}\left(\kappa\,{{\pi_{_K}}^2\over K}
+{\epsilon\over\kappa}\,K\right)
+{1\over2}\,\left({{\pi_{\phi}}^2\over K^3}
+{\mu}^2\,{\phi}^2\,K^3\right)
\equiv H_{_G}+H_{_M}
\ ,
\ee
where:
\begin{eqnarray}
\pi_{_K}&=&-{1\over\kappa}\,K\,\dot K \nonumber\\
& & \\
\pi_{\phi}&=&K^3\,\dot\phi 
\ .
\nonumber
\end{eqnarray}
Canonical quantisation
($\pi_{_K}\to\hat\pi_{_K}=-i\,\hbar\,\partial/\partial K$,
$\pi_\phi\to\hat\pi_\phi=-i\,\hbar\,\partial/\partial\phi$)
then leads to the Wheeler--DeWitt equation:
\be
{1\over 2}\,\left[{\kappa\,\hbar^2}\,{\partial_K^2\over K}
-{\epsilon\over\kappa}\,K
-{\hbar^2\over K^3}\,
{\partial^2\over\partial\phi^2}
+\mu^2\,\phi^2\,K^3
\right]\,\Psi(K,\phi)=0
\ .
\label{wdw}
\ee
One may now choose an {\em operator ordering\/} in the 
gravitational kinetic term given by:
\be
{\partial_K^2\over K}\equiv{\partial^2\over\partial K^2}\,{1\over K}
\ ,
\label{ordering}
\ee
and following a previously employed procedure \cite{brout}
(analogous to that used in molecular dynamics)
we express $\Psi$ in the factorized form:
\be
\Psi(K,\phi)=K\,\psi(K)\,\chi(\phi,K)
\ ,
\label{factor}
\ee
which after multiplying on the L.H.S. of Eq.~(\ref{wdw})
by $\chi^\ast$
and integrating over the matter degrees of freedom
leads to an equation for the gravitational part:
\begin{eqnarray}
& &{1\over2}\left[
\left({\kappa\,\hbar^2}\,{\partial^2\over\partial K^2}
-{\epsilon\over\kappa}\,K^2\right)
+{1\over\pro{\tilde\chi}{\tilde\chi}}\,
\bra{\tilde\chi}\,
\left({{\hat\pi_{\phi}}^2\over K^2}
+{\mu}^2\,\phi^2\,K^4\right)
\,\ket{\tilde\chi}\right]\,\tilde\psi
\equiv \nonumber\\   
& &\equiv\left[\hat H_{_G}\,K+K\,\expec{\hat H_{_M}}
\right]\,\tilde\psi 
=-{\kappa\,\hbar^2\over 2}\,
{\bra{\tilde\chi}\,\partial_K^2\,\ket{\tilde\chi}\over
\pro{\tilde\chi}{\tilde\chi}}\,\tilde\psi
\equiv-{\kappa\,\hbar^2\over 2}\,\expec{\partial_K^2}\,\tilde\psi
\ ,
\label{wdw_g}
\end{eqnarray}
where we have defined a scalar product:
\be
\pro{\chi}{\chi}\equiv
\int d\phi\,{\chi}^\ast(\phi,K)\,
\chi(\phi,K)
\ .
\ee
Further we have set:
\begin{eqnarray}
\psi&=&\exp\left\{-i\,\int^K A(K')\,dK'\right\}\,\tilde\psi
\nonumber \\
& & \label{fase}\\
\chi&=&\exp\left\{+i\,\int^K A(K')\,dK'\right\}\,\tilde\chi
\ ,
\nonumber
\end{eqnarray}
with:
\be
A\equiv -i\,
{\bra{\tilde\chi}\,\partial_K\,\ket{\tilde\chi}\over
\pro{\tilde\chi}{\tilde\chi}}
\ .
\ee
If we multiply Eq.~(\ref{wdw_g}) by $\tilde\chi$ and subtract it
from Eq.~(\ref{wdw}) we obtain:
\be
\tilde\psi\,K
\,\left[\hat H_{_M}-\expec{\hat H_{_M}}\right]\,\tilde\chi
+{\kappa\,\hbar^2}\,
\left({\partial\tilde\psi\over\partial K}\right)\,
{\partial\tilde\chi\over\partial K}
={\kappa\,\hbar^2\over2}\,\tilde\psi\,\left[
\expecl{{\partial^2\over\partial K^2}}-\partial_K^2\right]\,
\tilde\chi
\ ,
\label{wdw_m}
\ee
which is the equation for the matter (dust) wave function.
Let us emphasize that by a suitable choice of operator ordering
we have obtained a satisfactory form at the origin
($K=0$) for the gravitational part ($K\,\tilde\psi$) of the wave
function.
\subsection{The dust wave function}
\label{m_w}
If we consider the semiclassical (WKB) approximation to the wave 
function $\tilde\psi$ one has:
\be
{\partial\ln\tilde\psi\over\partial K}\simeq
-{i\over\hbar}\,{\partial S_{eff}\over\partial K}
=-{i\over\hbar}\,\pi_{_K}
\ ,
\label{wkb}
\ee
where $S_{eff}$ is the effective action satisfying
the Hamilton--Jacobi equation associated with the
L.H.S. of Eq.~(\ref{wdw_g}), that is:
\be
S_{eff}=-{1\over2\,\kappa}\,\int d\eta\,
\left[K^2\,\dot K^2-K\,
\left(\epsilon\,K-\expec{{\hat H}_{_M}}\right)
\right]
\ .
\label{s_eff}
\ee
In such a semiclassical limit $\tilde\psi$ will be 
peaked at $K=K_c$ (classical trajectory).
One may then define a (conformal) time variable:
\be
{\partial\over\partial\eta}\equiv
-i\,\hbar\,\kappa\,{\partial\ln\tilde\psi\over\partial K}\,
{\partial\over\partial K}
\ .
\label{eta}
\ee
Further if the R.H.S. of Eq.~(\ref{wdw_m}) is small: 
\be
{\kappa\,\hbar^2\over2}\,\left[
\expecl{{\partial^2\over\partial K^2}}
-\partial_K^2\right]\,\tilde\chi
\ll K_c\,\expec{\hat H_{_M}}
\ ,
\label{rhs_m}
\ee
one gets from Eq.~(\ref{wdw_m}) the Schr\"odinger 
equation for a harmonic oscillator:
\be
K_c\,\hat H_{_M}\,\chi_s=i\,\hbar\,
{\partial\chi_s\over\partial\eta}
\ ,
\label{schro}
\ee
where we have scaled the dynamical phase:
\be
\chi_s\equiv\tilde\chi\,\exp\left\{-{i\over\hbar}\,\int^\eta
\expec{\hat H_{_M}(\eta')}\,K_c\,d\eta'\right\} 
\ ,
\ee
and omitted $\tilde\psi$ while setting $K=K_c$ which is where the
semiclassical gravitational wave function is peaked and
is a solution to the classical equation of motion obtained
from Eq.~(\ref{s_eff}).
The solutions are \cite{venturi}:
\be
\tilde\chi=\tilde\chi_{_N}=N_{_N}\,H_{_N}(\gamma\,\phi)\,
e^{-\gamma^2\,\phi^2/2}=\pro{\phi}{\tilde\chi_{_N}}
\ ,
\label{chi_s}
\ee
where $H_{_N}$ is the Hermite polynomial of degree 
$N$, $N_{_N}$ a normalization factor
and:
\be
{\gamma}^2\equiv{\mu\over\hbar}\,{K_c}^3
\ .
\ee
Further:
\be
K_c\,\expec{\hat H_{_M}}_{_N}=
(N+1/2)\,{\hbar^2\,{\gamma}^2\over {K_c}^2}
=(N+1/2)\,\hbar\,\mu\,K_c
\equiv(N+1/2)\,\hbar\,\omega_c
\ .
\label{e_m}
\ee
\par
We may now consider the classical limit for the matter wave
function.
We introduce:
\begin{eqnarray}
\chi_{c}(\phi,\eta)& = &
\sqrt{\pi^{1/2}\over\gamma}\,\sum\limits_N\,
{(\gamma\,\phi_0)^N\over(2^N\,N!)^{1/2}}\,
e^{-\gamma^2\,{\phi_0}^2/4}\,
e^{-i\,(N+1/2)\,\omega_c\,(\eta-\eta_{_0})}\,
\tilde\chi_{_N}(\phi) 
\nonumber \\
& = &\sqrt{\pi^{1/2}\over\gamma}\,
\exp\left\{-{1\over2}\,{\gamma}^2\,\left[
\phi-\phi_0\,\cos\left(\omega_c\,(\eta-\eta_{_0})\right)
\right]^2-i\,\left[{1\over2}\omega_c\,(\eta-\eta_{_0})
\right.
\right.
\nonumber \\
& & \ \ \ \left.
\left.
+{\gamma}^2\,\phi_0\,\phi_i\,
\sin\left(\omega_c\,(\eta-\eta_{_0})\right)
-{1\over4}\,{\gamma}^2\,{\phi_0}^2\,
\sin\left(2\,\omega_c\,(\eta-\eta_{_0})\right)
\right]\right\}
\ .
\label{pac}
\end{eqnarray}
The superposition in Eq.~(\ref{pac}) corresponds to a minimum 
wave packet displaced
a distance $\phi_0$ at $\eta=\eta_{_0}$ and oscillating
about $\phi=0$ with a frequency $\omega_c$ which is 
constant in the adiabatic approximation.
In the classical limit, $\hbar\to 0$, one has $\gamma\to\infty$
and the wave function Eq.~(\ref{pac}) only has support on the
classical trajectory:
\be
|\chi_{cl}(\phi,\eta)|^2\sim
\delta\left(\phi-\phi_c(\eta)\right)
\ ,
\ee
with:
\be
\phi_c(\eta)=
\phi_0\,\cos\left(\omega_c\,(\eta-\eta_{_0})\right)
\ .
\label{osc_i}
\ee
The coefficients in Eq.~(\ref{pac}):
\be
{(\gamma\,\phi_0)^N\over(2^N\,N!)^{1/2}}\,
e^{-\gamma^2\,{\phi_0}^2/4}\equiv
{(N_c)^N\over\sqrt{N!}}\,
e^{-N_c/2}
\ ,
\ee
have a maximum at $N=N_c={\gamma}^2\,{\phi_0}^2/2$
and their square obeys a Poisson distribution with 
average value $N_c$ and standard variation $\sqrt{N_c}$.
Hence, if we just retain the term associated with the 
maximum in the dust wave function ($\chi_c\simeq\chi_{_{N_c}}$) 
one will have for the expectation value of the matter Hamiltonian
for such a state:
\be
K_c\,\expec{\hat H_{_M}}_c=
\sum\limits_{N=0}^\infty\,
{(N_c)^{2\,N}\over N!}\,e^{-N_c}\,
(N+1/2)\,\hbar\,\omega_c
\simeq
N_c\,\hbar\,\omega_c
\ ,
\label{h_m^c}
\ee
which is just the value of the classical Hamiltonian
$K_c\,H_{_M}$ for the oscillator in Eq.~(\ref{osc_i}):
\be
{1\over2}\left[{K_c}^2\,
\left({\partial\phi_c\over\partial\eta}\right)^2
+\mu^2\,{K_c}^4\,{\phi_c}^2\right]
={1\over2}\,\,{\gamma}^2\,{\phi_0}^2
\,\hbar\,\omega_c
\ .
\ee
\subsection{The gravitational wave function}
\label{s_wkb}
The existence of the semiclassical limit implies that
the gravitational wave function is peaked
around some classical trajectory $K_c$ whose Hamilton--Jacobi
action is given by $S_{eff}$ in Eq.~(\ref{s_eff}).
Since $\expec{\hat H_{_M}}$ is independent of $K_c$
the trajectory will be given by one of 
the $K_{cl}$ defined in Eq.~(\ref{hom}).
This implies that $S_{eff}$ must be equal to $S_{cl}=S_{_G}+S_{_M}$ 
given in Eqs.~(\ref{s_cl}). 
On comparing Eq.~(\ref{s_eff}) with Eq.~(\ref{s_cl}) above, one 
finds that in the L.H.S. of Eq.~(\ref{wdw_g}) one has:
\be
K_c\,\expec{\hat H_{_M}}=
K_{cl}\,\expec{\hat H_{_M}}={1\over\kappa}\,K_0\,K_{cl}
\ .
\label{h_m=dust}
\ee
Moreover the R.H.S. of Eq.~(\ref{wdw_g}) must be negligible: 
\be
-{\kappa\,\hbar^2\over2}\,\expecl{{\partial^2\over\partial K^2}}\ll
K_{cl}\,\expec{\hat H_{_M}}
\ .
\label{rhs_g}
\ee
Eq.~(\ref{h_m=dust}) means that the mean matter Hamiltonian
behaves as dust upon neglecting fluctuations (Eq.~(\ref{rhs_g})
and the R.H.S. of Eq.~(\ref{wdw_m})) and matter follows 
gravitation adiabatically (Born--Oppenheimer approximation).
\par
It is straightforward to derive an expression for $\tilde\psi$
which satisfies Eq.~(\ref{wdw_g}) in the semiclassical limit and
neglecting fluctuations.
Suppose that at time $\eta_{_0}$ for which 
$\partial_\eta K_{cl}=0$ (that is $\eta_{_0}=0$ or $\pi$
for $\epsilon=+1$ and $\eta_{_0}=0$ for $\epsilon=0,-1$) 
the gravitational wave function is a gaussian packet centred
at the value $K_{cl}(\eta_{_0})$ with width $b$:
\be
\tilde\psi(K,\eta_{_0})=
\exp\left\{-{\left(K-K_{cl}(\eta_{_0})\right)^2\over2\,b^2}\right\}
\ ,
\ee
where for simplicity we have omitted a normalization factor. 
At a succeeding time $\eta$ $\tilde\psi$ will be given by:
\be
\tilde\psi(K,\eta)=\int dK'\,
G_\epsilon(K-\epsilon\,K_{_0},\eta;K'-\epsilon\,K_{_0},\eta_{_0})\,
\tilde\psi(K';\eta_{_0})
\ ,
\ee
where $G_\epsilon$ is the Green's function for the (inverted) 
harmonic oscillator for $\epsilon=+1(-1)$ or for a particle moving 
in a linear potential for $\epsilon=0$ \cite{feynman}.
One then obtains:
\begin{eqnarray}
\tilde\psi(K,\eta)&=&\exp
\left\{i\,{(K-\epsilon\,K_{_0})^2\over2\,\hbar\,\kappa\,
(\partial^2_\eta h_\epsilon)}\,
(1-\alpha^2\,b^2)\,\sqrt{1-\epsilon\,
(\partial^2_\eta h_\epsilon)^2}\right\}
\nonumber\\
& &\times\,
\exp\left\{i\,{\alpha^2\,\hbar\,\kappa\over 2\,b^2}\,
K_{_0}\,(\partial^2_\eta h_\epsilon)\,\left(
K_{_0}\,\sqrt{1-\epsilon\,(\partial^2_\eta h_\epsilon)^2}
+2\,\epsilon\,(K-\epsilon\,K_{_0})\right)\right\}
\nonumber \\
& &\times\,
\exp\left\{-{\alpha^2\over2}\,\left(K-K_{cl}(\eta)\right)^2\right\}
\ ,
\label{psi}
\end{eqnarray}
where $h_\epsilon(\eta)$ has been defined in Eq.~(\ref{h}) and:
\be
\alpha={b\over\left[\kappa^2\,\hbar^2\,
(\partial^2_\eta h_\epsilon)^2+b^4\,
\left(1-\epsilon\,(\partial^2_\eta h_\epsilon)^2\right)
\right]^{1/2}} 
\ .
\label{alpha}
\ee
From Eq.~(\ref{psi}) one immediately obtains:
\be
{\partial\ln\tilde\psi\over\partial K}=
i\,{K_{_0}\over\hbar\,\kappa}\,\partial^2_\eta h_\epsilon(\eta)
+O(\hbar^0)
=-{i\over\hbar}\,\pi_{_K}+O(\hbar^0)
\ ,
\ee
in agreement with Eq.~(\ref{wkb}) to leading order in $\hbar$.
We further note that the total gravitational wave function 
is $K\,\tilde\psi(K)$ which is $0$ for $K=0$ as desired:
this solution is a consequence of our choice of ordering,
which however is not relevant in the semiclassical limit.
\par
The absolute value of the gravitational wave function is:
\be
|\tilde\psi(K,\eta)|^2=\exp\left\{-\alpha^2\,
\left(K-K_{cl}(\eta)\right)^2\right\}
\ ,
\ee
If one now considers the limit $\hbar\to 0$ followed by $b\to 0$
(the {\em classical point--like limit\/})
the classical trajectory is obtained:
\be
\alpha\to\infty\ \ \ \Longrightarrow\ \ \ 
{K^2\,|\tilde\psi|^2\over\bra{\tilde\psi}\,\hat K^2\,\ket{\tilde\psi}}
\to\delta(K-K_{cl})
\ ,
\label{hp=0}
\ee
where $\bra{\tilde\psi}\,\hat K^2\,\ket{\tilde\psi}$ 
is the norm of the complete wave function
$K\,\tilde\psi$ and is given in Appendix~\ref{norme}.
With the above limits interchanged, that is $b\to 0$ 
with $\hbar$ finite, one would obtain:
\be
\alpha\to 0\ \ \ \Longrightarrow\ \ \ 
{K^2\,|\tilde\psi|^2\over\bra{\tilde\psi}\,\hat K^2\,\ket{\tilde\psi}}
\to 0
\ .
\label{b=0}
\ee
However, we expect that when $b$ becomes smaller than the Planck 
length $\ell_p$ quantum gravitational effects (fluctuations)
become significant.
Hence it is more sensible to consider the limit for which one
has $b\sim\ell_p$ (corresponding to a minimum size wave packet
of the order of the Planck length)
and then consider $\ell_p\to 0$.
For such a case:
\be
\alpha\sim {\ell_p}^{-1}\to\infty
\ ,
\label{l=0}
\ee
which again leads to Eq.~(\ref{hp=0}).
\par
If we also consider the semiclassical limit for matter from
Eqs.~(\ref{pac}) and (\ref{h_m=dust}) one has:
\be
{1\over\kappa}\,K_0\,K_{cl}=\,N_c\,\hbar\,\mu\,K_{cl}
\ ,
\ee
and, taking into account the junction condition Eq.~(\ref{M}):
\be
M_{_G}={\rho_{_0}}^3\,\,N_c\,\hbar\,\mu
\ .
\label{m=sum}
\ee
We may then conclude that the total mass $M_{_G}$ 
of the collapsing sphere is proportional to the number
(large) of dust quanta $N_c$.
The total euclidean action $S_e$ (the gravitational entropy)
of the system once the black hole has formed is known to be 
proportional to the area of the horizon \cite{gibbons}. 
On using our result Eq.~(\ref{m=sum}) it also follows that:
\be
S_e={A\over 4}=4\,\pi\,M^2=4\,\pi\,\left(\,{\rho_{_0}}^3\,
{\ell_p}^2\,N_c\,\mu\right)^2
\ ,
\ee
which implies that $S_e$ is also quantised in terms of the dust 
quanta.
\par
The above apparently differ from the result obtained previously,
$M_{_G}\sim m_p\sqrt{n}$ \cite{bekenstein}, 
but we observe that the $N_c$ appearing in Eq.~(\ref{m=sum})
are energy quantum numbers of the dust, while the $n$ 
in Bekenstein's formula comes from quantising 
the gravitational Hamiltonian.
In fact, let us consider the case $\epsilon=+1$.
Eq.~(\ref{wdw_g}), using Eqs.~(\ref{h_m=dust}) and (\ref{rhs_g}),
becomes:
\be
{1\over 2}\,\left[-\kappa\,\hbar^2\,{\partial^2\over\partial K^2}
+{K\over\kappa}\,\left(K-2\,K_{_0}\right)
\right]\,\tilde\psi=0
\ ,
\ee
which may be re--written as 
\cite{peleg}:
\be
{1\over2}\left[{\kappa\,\hbar^2}\,
{\partial^2\over\partial {X_n}^2}
+{1\over\kappa}\,{X_n}^2
\right]\,\tilde\psi=
{{K_{0,n}}^2\over2\,\kappa}\,\tilde\psi
\ ,
\ee
where $X_n\equiv K-K_{0,n}$.
The solutions are:
\be
\tilde\psi=H_n(X_n/\ell_p)\,e^{-{X_n}^2/2\,\ell_p}
\ ,
\ee
with $H_n$ the Hermite polynomial of degree $n$ 
and ${K_{0,n}}^2={\ell_p}^2\,(n+1/2)$.
It then follows that:
\be
M_{_G}=m_p\,{\rho_{_0}}^3\,\sqrt{n+1/2}
\ ,
\ee
which is essentially Bekenstein's formula.
Let us note that in the other two cases $\epsilon=0,-1$
no analogous discrete quantisation exists.
\subsection{Consistency conditions}
\label{test}
Let us first see how quantisation modifies the geodesic
motion since, even if $\tilde\psi(K,\eta)$ is peaked on the classical
value $K_{cl}$, one has fluctuations around it.
Ehrenfest's theorem \cite{greensite} (with the classical limit)
will be satisfied if the average
position $\expec{\hat r}$ will coincide with the
classical trajectory up to  small fluctuations.
That is:
\be
\rho_{_0}\,{\bra{\tilde\psi}\hat K^3\ket{\tilde\psi}
\over\bra{\tilde\psi}\,\hat K^2\,\ket{\tilde\psi}}
\equiv\rho_{_0}\,\expec{\hat K}
=r_{_0}(\eta)+O(\hbar)
\ .
\label{ehrenfest}
\ee
What one actually obtains is (see Appendix~\ref{norme} 
for the details):
\be
\expec{\hat K}=
{\sqrt{\pi}\,\alpha\,K_{cl}\,
\left(3+2\,\alpha^2\,{K_{cl}}^2\right)\,
\mbox{\rm Erfc}(-\alpha\,K_{cl})
+2\,\left(1+\alpha^2\,{K_{cl}}^2\right)\,e^{-\alpha^2\,{K_{cl}}^2}
\over
\alpha\,\left[\sqrt{\pi}\,\left(1+2\,\alpha^2\,{K_{cl}}^2\right)\,
\mbox{\rm Erfc}(-\alpha\,K_{cl})
+2\,\alpha\,K_{cl}\,e^{-\alpha^2\,{K_{cl}}^2}\right]}
\ ,
\ee
and in the classical point--like limit, Eq.~(\ref{hp=0})
or Eq.~(\ref{l=0}),
this expression gives exactly the classical trajectory:
\be
\rho_{_0}\,\expec{\hat K}
\to\rho_{_0}\,K_{cl}
\ .
\ee
\par
Ehrenfest's theorem, as illustrated in Eq.~(\ref{ehrenfest}),
implies that fluctuations are negligible.
In our case this means that one must have:
\be
\Delta\equiv{|\expec{\hat K^2}-\expec{\hat K}^2|
\over \expec{\hat K}}\ll\expec{\hat K}
\ .
\label{delta}
\ee
For the wave function $\tilde\psi$ one finds
(see Appendix~\ref{norme}):
\be
\begin{array}{c}
\expec{\hat K^2}\equiv\strut\displaystyle
{\bra{\tilde\psi}\,\hat K^4\,\ket{\tilde\psi}\over
\bra{\tilde\psi}\,\hat K^2\,\ket{\tilde\psi}}=\\
\strut\displaystyle
{\sqrt{\pi}\,\left(3+12\,\alpha^2\,{K_{cl}}^2
+4\,\alpha^4\,{K_{cl}}^4\right)\,
\mbox{\rm Erfc}(-\alpha\,K_{cl})
+2\,\alpha\,{K_{cl}}\,
\left(5+2\,\alpha^2\,{K_{cl}}^2\right)
\,e^{-\alpha^2\,{K_{cl}}^2}
\over  2\,\alpha^2\,\left[
\sqrt{\pi}\,\left(1+2\,\alpha^2\,{K_{cl}}^2\right)\,
\mbox{\rm Erfc}(-\alpha\,K_{cl})
+2\,\alpha\,K_{cl}\,e^{-\alpha^2\,{K_{cl}}^2}
\right]}
\ ,
\end{array}
\ee
so that:
\be
{\Delta\over\expec{\hat K}}\sim{1\over\alpha^2}
\to 0
\ ,
\ee
in the classical point--like limit, Eq.~(\ref{hp=0})
or Eq.~(\ref{l=0}).
However if one considers the limiting procedure used in
Eq.~(\ref{b=0}) one has:
\be
{\Delta\over\expec{\hat K}}\to {3\,\pi\over8}-1\simeq 0.18
\ ,
\ee
which implies Eq.~(\ref{ehrenfest}) is not satisfied.
\par
It is also interesting to consider small black holes.
Let us then take $K_{cl}$ small at fixed $\hbar$ and 
$b\sim\ell_p$, one then has: 
\be
\Delta\sim {3\,\pi-8\over4\,\sqrt{\pi}}\,{1\over\alpha}\sim
\ell_p
\ ,
\ee
which of course implies that one must have
$r_{_0}(\eta)=\rho_{_0}\,K_{cl}(\eta)\gg\ell_p$.
Since we are only interested in the limit for which
$r_{_0}$ approaches $r_{_H}$ from outside, one must have 
$2\,M\gg\Delta\sim\ell_p$ or:
\be
M_{_G}\gg m_p
\ .
\label{cond}
\ee
This coincides with what one would expect from 
naive quantum mechanical considerations.
Indeed, if $\lambda_{bh}=\kappa\,\hbar/M$ is the Compton
wave length of the black hole, it should be much less
than its Schwarzschild radius $2\,M$ for the semiclassical
approximation to work.
This corresponds exactly to Eq.~(\ref{cond}).
\par
The remaining consistency conditions are Eqs.~(\ref{rhs_g})
and (\ref{rhs_m})
to which one may add the condition for the validity of the
adiabatic approximation, that is that the time dependence in
the matter wave function induced by the slow variable
(gravitation) be small.
This essentially requires that in Eq.~(\ref{wdw_m}) the second term on
the L.H.S. be larger than the third (evaluated using 
Eq.~(\ref{wkb})).
On using the the solutions Eq.~(\ref{chi_s}) for $N=N_c$
which will approximate classical dust
(analogous results are obtained for the ``vacuum'' state
$N=0$ for matter) 
into the R.H.S. of Eq.~(\ref{wdw_g}) one obtains
(see Appendix~\ref{flutt} for more details):
\be
{\kappa\,\hbar^2\over 2}\,\expecl{{\partial^2\over\partial K^2}}
\simeq
{9\,\kappa\,\hbar^2\over 16\,{K_{cl}}^2}
\,\left({N_c}^2+N_c+1\right)
\le
{9\,\kappa\,\hbar^2\,{\rho_{_0}}^2\over 64\,M^2}
\,\left({N_c}^2+N_c+1\right)
\ ,
\ee
where $K_{cl}$ is evaluated at the minimum value we need, 
that is $r_{_H}/\rho_{_0}=2\,M/\rho_{_0}$.
The condition Eq.~(\ref{rhs_g}) then becomes:
\begin{eqnarray}
{\kappa\,\hbar^2\over 2}\,{\expec{\partial_K^2}\over
K_{cl}\,\expec{\hat H_{_M}}}
&\simeq&
{9\,\kappa\,\hbar\over 8\,\mu\,{K_{cl}}^3}\,
\,{{N_c}^2+{N_c}+1\over 2\,{N_c}+1} 
\nonumber \\
&\le&
{9\,\kappa\,\hbar\,{\rho_{_0}}^3\over 64\,\mu\,M^3}\,
\,{{N_c}^2+{N_c}+1\over 2\,{N_c}+1}
\nonumber \\ 
&\ll& 1
\ ,
\label{cond_1}
\end{eqnarray}
Analogously, Eq.~(\ref{rhs_m}) yields:
\begin{eqnarray}
{\kappa\,\hbar^2\over2}\,{\left|\left[
\expec{\partial_K^2}-\partial^2_K\right]\,\tilde\chi\right|
\over K_{cl}\,\expec{\hat H_{_M}}}
&\simeq&
{\kappa\,\hbar\,\over \mu\,{K_{cl}}^3}\,
\,{{N_c}^2+{N_c}+1\over 2\,{N_c}+1} 
\nonumber \\
&\le&
{\kappa\,\hbar\,{\rho_{_o}}^3\over 8\,\mu\,M^3}\,
\,{{N_c}^2+{N_c}+1\over 2\,{N_c}+1}
\nonumber \\ 
&\ll& 1
\ ,
\label{cond_2}
\end{eqnarray}
which is essentially the same as Eq.~(\ref{cond_1}).
\par
Finally from Eqs.~(\ref{wdw_m}) and (\ref{eta}) one must 
have:
\begin{eqnarray}
\hbar\,{\partial K_{cl}\over\partial\eta}\,
{\left|\expec{\stackrel{\leftarrow}{\partial_K}
\,\partial_K}\right|^{1/2}
\over K_{cl}\,\expec{\hat H_{_M}}}
&\simeq&
{\partial K_{cl}\over\partial\eta}\,
{\sqrt{{N_c}^2+N_c+1}\over\mu\,{K_{cl}}^2\,(2\,N_c+1)}
\nonumber \\
&\le&
{3\over2\,\sqrt{2}}\,
{1\over\mu^2\, M}\,
{\sqrt{{N_c}^2+N_c+1}\over(2\,N_c+1)}\times
\left\{\begin{array}{ll}
\sqrt{1-{\rho_{_0}}^2} & \epsilon=+1\\
1 &\epsilon=0\\
\sqrt{1+{\rho_{_0}}^2} & \epsilon=-1
\end{array}\right.
\nonumber \\
&\ll& 1
\ ,
\label{cond_3}
\end{eqnarray}
which, for the periodic case $\epsilon=+1$, can be understood
as the ratio between the periods of the gravitational motion
and the one for the matter state $\chi_{_{N_c}}$;
the fact it is small is a statement of the adiabatic
approximation.
Since for the semiclassical wave packet in Eq.~(\ref{pac}) 
$N_c\gg 1$, the above three conditions are essentially the same and,
on using Eq.~(\ref{m=sum}) to express $N_c$ as a function
of $M$, one obtains:
\be
M\gg {1\over\mu}
\ ,
\label{cond_m}
\ee
which means that the Schwarzschild radius of the dust must
be much larger than the Compton wave length of the dust
particles, or: 
\be
M_{_G}\gg {{m_p}^2\over m}
\ .
\label{cond_0}
\ee
Since $m\ll m_p$ for all known elementary particles, 
the latter expression is a stronger condition than 
Eq.~(\ref{cond}).
The fact that the adiabatic approximation is not valid
if Eq.~(\ref{cond_m}) is not satisfied is not surprising.
Indeed in such a case fluctuations, corresponding
to the creation of matter particles, are large
(and the evaporation time small \cite{hawking}).
\setcounter{equation}{0}
\section{Coupling with a conformal scalar field}
\label{radiation}
An observer modelled by a point--like monopole detector
localized at a fixed $r>r_{_H}$ and coupled to the radiation 
field can be used to reveal Hawking radiation \cite{hawking}.
Since the detector trajectory approaches the uniformly accelerated
one in Minkowski space--time as $r\to r_{_H}$, one may take the 
radiation field in the Unruh vacuum and proceed as
in the calculation of the Unruh effect \cite{unruh}.
One then obtains that the observer detects
a thermal flux with a temperature given by:
\be
T_{_H}={\hbar\over 8\,\pi\,M}
\ .
\ee
For the flat case (Minkowski space--time) one can also consider 
a quantum detector described by a gaussian wave packet and
one gets the usual Unruh effect
for a classical point--like limit as in Eq.~(\ref{hp=0})
\cite{casadio}.
\par
Here we shall examine a different approach.
Let us consider the outer dust shell of the sphere situated 
at $\rho=\rho_{_0}$
as a quantum system characterized by an internal energy spectrum
which is related to the number of dust quanta
in the shell itself.
From the point of view of a distant observer
the position of the shell is not a classical variable but is 
a quantum observable determined by the gravitational wave function 
describing the semiclassical collapse obtained in the previous 
section.
\par
We consider an isotropic massless scalar 
field $\varphi=\varphi(\rho,\eta)$ conformally coupled
to gravity \cite{birrell} and to the outer shell of dust.
Its Lagrangian density will be given by:
\begin{eqnarray}
{\cal L}_\varphi&=&-{1\over 2}\,\left[
\partial_\mu\varphi\,\partial^\mu\varphi
+{1\over 6}\,R\,\varphi^2\right]
\nonumber \\
& &+\int d\tau\,\int d\rho\,\int_0^{+\infty} dK\,
{|K\tilde\psi(K,\tau)|^2\over\bra{\tilde\psi}\,\hat K^2\,\ket{\tilde\psi}}\,
\delta(\rho-\rho_{_0})\,
Q(\rho,\tau)\,\varphi(\rho,\tau)
\ ,
\end{eqnarray}
where $R$ is the curvature scalar, $Q(\rho,\tau)$ describes a particle 
(monopole) emitter and the factor $\delta(\rho-\rho_{_0})\,
|K\tilde\psi|^2/\bra{\tilde\psi}\,\hat K^2\,\ket{\tilde\psi}$
forces the interaction to be localized on the outer shell.
In particular, in the classical point--like limit
(Eq.~(\ref{hp=0})) it leads to $\delta(\rho-\rho_{_0})\delta(K-K_{cl})
=\delta(r-r_{_0})$.
The reason we consider an excited monopole (emitter) is to avoid
the problem of back--reaction on matter and gravity.
Further in our case the observer is distant from the dust
sphere and static, this is the reason we shall use
Schwarzschild coordinates for it.
\par
Let us assume the emitter has a discrete set of internal
energy eigenstates described by $\ket{E}$
where $E$ is the energy observed by our distant observer.
We suppose the emitter is initially in an excited state
$E_{_0}$ and decays by emitting quanta of the scalar field
$\varphi$ to a state $E_{_0}-\hbar\,\omega$.
One may then estimate, using first order perturbation
theory, the total probability amplitude $P(\omega,\bar\eta)$
for the emitter to decay in a finite conformal time.
It will be given by:
\begin{eqnarray}
P(\omega,\bar\eta)&=&{{Q_\omega}^2\over\hbar^2}\,
\int_{\tau(\eta_{_0})}^{\tau(\bar\eta)} d\tau''\,
\int_{\tau(\eta_{_0})}^{\tau(\bar\eta)} d\tau'\,
e^{i\,\omega\,(t''-t')}\,
D_\epsilon^+(\eta'',\eta')
\nonumber \\
&=&
{{Q_\omega}^2\over\hbar^2}\,
\int^{\bar\eta}_{\eta_{_0}} K_{cl}(\eta'')\,d\eta''\,
\int^{\bar\eta}_{\eta_{_0}} K_{cl}(\eta')\,d\eta'
\,e^{i\,\omega\,(t''-t')}\,D^+_\epsilon(\eta'',\eta')
\ ,
\label{p}
\end{eqnarray}
where
$Q_\omega\equiv|\bra{E_0}\,Q(\rho_{_0},0)\,\ket{E_0-\hbar\,\omega}|$
is the absolute value of the matrix
element of the monopole between the states with 
energy $E_0$ and $E_0-\hbar\,\omega$,
$t''=t(\eta'')$ and 
$t'=t(\eta')$ are the Schwarzschild times measured by the distant
observer expressed in term
of the conformal times as given in Eq.~(\ref{t(eta)}),
and $\bar\eta$ is an upper cut off such that 
$\eta_{_H}<\bar\eta<\eta_{_0}$.
Further during the interval $(\eta_{_0},\bar\eta)$ the static
monopole emitter near the horizon is immersed in the gravitational
wave packet associated with the last dust shell and
$D^+_\epsilon$ is the Wightman function \cite{birrell}
for the isotropic conformal scalar field in the metric 
Eq.~(\ref{g_hom}) evaluated at the same spatial point 
$\rho''=\rho'=\rho$, but with the $K_{cl}$ in the 
denominator ``smeared'' by the wave function $K\tilde\psi$ 
of Eq.~(\ref{psi}):
\begin{eqnarray}
D^+_\epsilon(\eta'',\eta')&\equiv&
\int_0^{+\infty} dK''\,
{|K''\tilde\psi(K'',\eta'')|^2\over
\bra{\tilde\psi(\eta'')}\,\hat K^2\,\ket{\tilde\psi(\eta'')}}
\nonumber \\
& &\times 
\int_0^{+\infty} dK'\,
{|K'\tilde\psi(K',\eta')|^2\over
\bra{\tilde\psi(\eta')}\,\hat K^2\,\ket{\tilde\psi(\eta')}}
\,{D_\epsilon(\eta'',\eta')\over K''\,K'} 
\ ,
\end{eqnarray}
where:
\be
D_\epsilon(\eta'',\eta')=\left\{
\begin{array}{ll}
\strut\displaystyle{\hbar\over 8\,\pi^2\,
\left[\cos(\eta''-\eta'-i\,\varepsilon)-1\right]}
&\ \ \ \epsilon=+1 \\
& \\
\strut\displaystyle-{\hbar\over 4\,\pi^2\,
(\eta''-\eta'-i\,\varepsilon)^2}
&\ \ \ \epsilon=0,-1 
\ .
\end{array}\right.
\ee
In using these propagators one is limiting oneself
to $s$-waves.
However this is known not to be too restrictive since there
is a potential barrier outside the horizon in the
Klein--Gordon equation for the scalar field which, 
in practice, suppresses all higher angular momentum modes.
On performing the $K''$, $K'$ integrations one finds:
\be
K_{cl}(\eta'')\,K_{cl}(\eta')\,D^+_\epsilon(\eta'',\eta')=
R(\eta'')\,R(\eta')\,D_\epsilon(\eta'',\eta')
\ ,
\label{w_f}
\ee
where (see Appendix~\ref{norme}):
\begin{eqnarray}
R(\eta)&\equiv&K_{cl}(\eta)\,
{\bra{\tilde\psi(\eta)}\,\hat K\,\ket{\tilde\psi(\eta)}
\over\bra{\tilde\psi(\eta)}\,\hat K^2\,\ket{\tilde\psi(\eta)}}
\nonumber \\
&=&2\,\alpha\,K_{cl}\,{\sqrt{\pi}\,\alpha\,K_{cl}\,
\mbox{\rm Erfc}(-\alpha\,K_{cl})+e^{-\alpha^2\,{K_{cl}^2}}
\over
\sqrt{\pi}(1+2\,\alpha^2\,{K_{cl}}^2)\,
\mbox{\rm Erfc}(-\alpha\,K_{cl})
+2\,\alpha\,K_{cl}\,e^{-\alpha^2\,{K_{cl}^2}}}
\ .
\label{r}
\end{eqnarray}
\par
It is now convenient to change the variable of integration
from the conformal time to the Schwarzschild time according 
to Eq.~(\ref{t(eta)}) and define:
\be
\left\{\begin{array}{l}
\Delta t\equiv t''-t' \\
\\
T\equiv(t''+t')/2
\ .
\end{array}\right.
\ee
Eq.~(\ref{p}), with $\eta''(t''=L)=\eta'(t'=L)=\bar\eta$
and $\eta''(t''=0)=\eta'(t'=0)=\eta_{_0}$,
then becomes:
\begin{eqnarray}
P(\omega,L)&\simeq&{{Q_\omega}^2\over\hbar^2}\,
\int_0^L dT\,{{C_\epsilon}^2\,e^{-{T\over M}}\over4\,M^2}\,
\int\limits^{L-2|T-L/2|}_{2|T-L/2|-L} d(\Delta t)\,
e^{i\,\omega\,\Delta t}
\nonumber \\
& &\times\,
R(T+\Delta t/2)\,R(T-\Delta t/2)\, 
D_\epsilon(\Delta t,T)
\ .
\label{p1}
\end{eqnarray}
We now note that $D_\epsilon$ is singular for:
\be
\begin{array}{ll}
\cos(\eta''-\eta'-i\,\varepsilon)=1 & \epsilon=+1 \\
& \\ 
(\eta''-\eta'-i\,\varepsilon)^2=0 & \epsilon=0,-1
\ .
\end{array}
\label{0}
\ee
Using Eq.~(\ref{eta(t)}) one has near the horizon:
\be
\eta''-\eta'\simeq
C_\epsilon\,e^{-{t''\over2\,M}}-C_\epsilon\,e^{-{t'\over2\,M}}
=-2\,C_\epsilon\,e^{-{T\over2\,M}}\,
\sinh\left({\Delta t\over4\,M}\right)
\ ,
\label{e''-e'}
\ee
hence on substituting into Eqs.~(\ref{0}) 
in the limit $\varepsilon\to 0$, one finds double poles at:
\be
\Delta t_q=4\,\pi\,i\,q\,M
\ ,
\label{poles}
\ee
where $q$ is an integer or zero.
Apart from these singularities the integrand in 
Eq.~(\ref{p1}) is an analytic function in the complex 
variable $\Delta t$.
Thus one can close the integration path 
with a semicircle of radius $L-2\,|T-L/2|$ in the
upper half complex $\Delta t$ plane and apply the theorem
of residues.
In the limit for $L\to\infty$ the integral along the 
semicircle vanishes due to the energy exponential
and the possible contributions come 
from the residues alone.
\par
Let us further examine the above singularities.
In terms of the conformal time near the horizon 
one finds:
\be
\begin{array}{l}
\eta''-\eta_{_H}=(-1)^q\,C_\epsilon\,e^{-{T\over2\,M}} \\
\\
\eta'-\eta_{_H}=(-1)^q\,C_\epsilon\,e^{-{T\over2\,M}}
\ .
\end{array}
\label{e_e}
\ee
It is clear that Eqs.~(\ref{e_e}) have the opposite sign to
the one in Eq.~(\ref{eta(t)}) for $q=2\,n+1$
and therefore we must consider only the residues at even $q$,
that is:
\be
\Delta t_{2\,n}=8\,\pi\,i\,n\,M 
\ .
\label{poli}
\ee
which are the poles that are associated with 
a free fall from outside the horizon.
The ones with odd $q$ correspond to geodesic motion leaving the
horizon from {\em inside\/}.
Thus for an observer outside the horizon the only relevant
poles are the ones given in Eq.~(\ref{poli}).
\par
The residues to be summed over then are:
\begin{eqnarray}
S_n&=&\lim\limits_{\Delta t\to\Delta t_{2n}}\,
{d\over d(\Delta t)}\left[(\Delta t-\Delta t_{2n})^2\,
e^{i\,\omega\,\Delta t}\,\bar D_\epsilon(\Delta t, T)\right]
\nonumber \\
&=&
-{4\,M^2\,\hbar\over{C_\epsilon}^2}
\,e^{T\over M}\,
{d\over d(\Delta t)}\left[
R(T+\Delta t/2)\,R(T-\Delta t/2)
e^{i\,\omega\,\Delta t}
\right]_{\Delta t=\Delta t_{2n}}
\ ,
\label{res}
\end{eqnarray}
where $R$ is given in terms of $\eta$ in Eq.~(\ref{r}).
Further:
\be
\left.{d\over d(\Delta t)}R(T+\Delta t/2)
\right|_{\Delta t=\Delta t_{2n}}=
-\left.{d\over d(\Delta t)}R(T-\Delta t/2)
\right|_{\Delta t=\Delta t_{2n}}
\ ,
\ee
so that the expression Eq.~(\ref{res}) simplifies to: 
\be
S_n=-{4\,M^2\,\hbar\over{C_\epsilon}^2}
\,e^{T\over M}\,R^2(T)\,i\,\omega\,e^{-\beta\,\hbar\,\omega\,n}
\ ,
\ee
with $\beta\equiv(8\,\pi\,M)/\hbar$.
\par
After performing the $\Delta t$ integration 
the probability amplitude becomes:
\begin{eqnarray}
P(\omega,L)&\simeq&
{{Q_\omega}^2\,\omega\over 2\,\pi\,\hbar}
\,\int_0^L dT\,R^2(T)\,
\sum\limits_{n=0}^N\,e^{-\beta\,\hbar\,\omega\,n} 
\ .
\label{p3}
\end{eqnarray}
where $N$ is the largest integer $\le(L-2\,|T-L/2|)/8\,\pi\,M$ 
and we have omitted the integral along the contour which does
not contribute for $L$ large.
We observe that in order to include at least
one pole inside the contour of integration
$L$ must be greater than $|\Delta t_2|=8\,\pi\,M$,
which implies:
\be
\eta-\eta_{_H}\simeq C_\epsilon\,e^{-{T\over 2\,M}}
<C_\epsilon\,e^{-2\,\pi}
\ ,
\label{sup}
\ee
and if we consider the shell very close to the horizon, 
that is values of $T$ large in Eqs.~(\ref{p3}) and (\ref{sup}),
we obtain a probability amplitude per unit time equal to:
\be
{P(\omega,L)\over L}\simeq{{Q_\omega}^2\,R^2(\infty)
\over 2\,\pi\,\hbar}\,{\omega\over 1-e^{-\beta\,\hbar\,\omega}}
\ ,
\label{p4}
\ee
which is a Planck distribution with the usual Hawking
temperature $T_{_H}=1/\beta$ and
$R(\infty)$ is a non zero constant given by $R(\eta_{_H})$.
Let us note that the terms omitted in the approximations that lead 
to Eq.~(\ref{p4}) are mainly concerned with 
effects at small time $T$:
in fact we let $L$ go to infinity and neglected factors
of $e^{-T/2\,M}$.
The corrections one expects when relaxing
these assumptions are relevant for small values of 
$T$ and should lead to transient effects
when the monopole starts radiating or the horizon
is forming.
\par
In the classical point--like limit, Eq.~(\ref{hp=0}),
$R\to 1$ and one recovers the usual
field theory in the fixed Robertson--Walker--like background.
In this last case Eq.~(\ref{p4}) in the large $L$ limit 
becomes:
\be
{P(\omega,L)\over L}\simeq{{Q_\omega}^2
\over 2\,\pi\,\hbar}\,{\omega\over 1-e^{-\beta\,\hbar\,\omega}}
\ ,
\label{p5}
\ee
which is equal to Eq.~(\ref{p4}) apart from the factor
of $R^2(\infty)$.
If we However take the alternative limit, Eq.~(\ref{b=0}),
that is $b\to0$,
one finds that the dust shell decouples
from the conformal field, since: 
\be
R\sim\alpha\to 0
\ ,
\ee
and there is no emission:
\be
P(\omega,L)\to 0
\ .
\label{p=0}
\ee
One may speculate that this effect can be used to eliminate
ultra--planckian effects.
In fact, it is known that, if $\omega$ is the frequency
of the emitted quanta as is measured by a distant
observer, a fixed observer located near the point of emission
at $r=r_{_0}$ will measure instead a blue--shifted frequency:
\be
\omega^\ast=\left(1-{2\,M\over r_{_0}}\right)^{-1/2}\,\omega
\ ,
\ee
and this expression clearly gives 
$\omega^\ast>m_p/\hbar={\ell_p}^{-1}$
for $r_{_0}$ sufficiently close to $r_{_H}$.
In order to probe these modes, one must use an emitter
(detector) localized in a region smaller than 
$\omega^{-1}\sim\ell_p$. 
So one expects that $b$ in the wave function 
Eq.~(\ref{psi}) should be less than $\ell_p$
for our collapsing shell to couple with conformal 
quanta of ultra--planckian energies.
But this would correspond to the limit in Eq.~(\ref{b=0})
which in turn implies Eq.~(\ref{p=0}):
ultra--planckian energy quanta are not emitted
by the monopole.
\par
A further point worth noting is that $R(\infty)$ 
given by $R(\eta=\eta_{_H})$ in Eq.~(\ref{r}) actually
depends on $\epsilon$ through $\alpha$ (see Eq.~(\ref{alpha}))
but not through $K_{cl}(\eta_{_H})=2\,M/\rho_{_0}$.
This of course implies that the probability amplitude
in Eq.~(\ref{p4}) depends on the internal dust sphere
metric.
Indeed since:
\be
\alpha(\eta_{_H})={b\over\left[
4\,{\ell_p}^4\,{\rho_{_0}}^2\,\left(1-\epsilon\,{\rho_{_0}}^2\right)
+b^4\,\left(1-2\,\epsilon\,{\rho_{_0}}^2\right)^2\right]^{1/2}}
\ ,
\ee
one would have in the limit $b\gg\ell_p$,
$\alpha(\eta_{_H})\,K_{cl}(\eta_{_H})\to 0$:
\be
R(\eta_{_H})\simeq
{2\over\sqrt{\pi}}\,\alpha(\eta_{_H})\,K_{cl}(\eta_{_H})
\simeq\left\{\begin{array}{ll}
\strut\displaystyle{4\,M\over\sqrt{\pi}}\,
{1\over b\,\rho_{_0}\,(2\,{\rho_{_0}}^2-1)}
&\ \ \ \ \ \ \ \ \epsilon=+1 \\
 & \\
\strut\displaystyle{4\,M\over\sqrt{\pi}}\,
{1\over b\,\rho_{_0}}
&\ \ \ \ \ \ \ \ \epsilon=0 \\
 & \\
\strut\displaystyle{4\,M\over\sqrt{\pi}}\,
{1\over b\,\rho_{_0}\,(2\,{\rho_{_0}}^2+1)} 
&\ \ \ \ \ \ \ \ \epsilon=-1 
\ ,
\label{stima}
\end{array}\right.
\ee
that is one has different emission rates depending on
the internal geometry of the sphere of dust.
\par
Let us end with a speculation:
we considered the matching condition in the classical limit,
if one wished to consider quantum mechanical corrections
one should replace it by the continuity of the gravitational
wave function inside the dust (Robertson--Walker)
with that outside (Schwarzschild).
This suggests the results in Eq.~(\ref{stima}) should not
change dramatically immediately outside the dust,
implying a form of quantum hair leading to information
on the geometry in the black hole through the intensity
of the radiation.
\setcounter{equation}{0}
\section{Conclusions}
We have studied the collapse of a self--gravitating
homogeneous sphere of dust in the context of canonical
general relativity by applying the Born--Oppenheimer approach
to the coupled dust-gravity system.
Such an approach allows for the quantization both of matter and
gravity and it is only in the semiclassical limit for the latter
and in the absence of fluctuations that time emerges and one
recovers quantum theory for matter on a classical background.
This has allowed us, in some measure, to estimate quantum
mechanical effects both for matter and gravity.
\par
The gravitational wave packets obtained correspond to the
quantum mechanical evolution of a gaussian packet of initial
width $b$ ($=\ell_p$ for a minimum wave packet) centred on
the classical trajectory $K_{cl}$.
This has allowed us to examine the conditions for which one
recovers the usual ``point--like'' classical collapse which, for
example, was necessary to implement the matching (or junction)
condition.
\par
Further the fact matter (dust) is also quantised has allowed us
to also estimate matter fluctuations which lead to a violation of
the adiabatic approximation.
The validity of the approximation not surprisingly requires that
the Schwarzschild radius of the dust sphere be greater than the
Compton wavelength of the matter (dust) particles, indeed should
this not be the case one would naturally
expect fluctuations corresponding to the creation of particles
by the gravitational field to be important.
\par
One of the most interesting quantum effects
in black hole physics is Hawking radiation and in order to 
to study this we considered a static monopole, with an internal
energy spectrum, situated near the horizon which is being formed
in coincidence with the collapse of the last shell.
The monopole couples gravitation to a conformal scalar field which
has support both inside and outside the dust sphere.
We found that a thermal Wightman's function corresponding to
a Planckian distribution with the Hawking temperature appeared
in the de-excitation (or excitation) probability of the monopole.
\par
Again the fact that our last dust shell is described by a wave
function allowed us to discuss various limits and differing orders
of limits.
Let us just mention that although one always had a thermal
Hawking temperature spectrum, the intensity of the radiation
depended on the internal dust metric.
Should the result also be true just outside the dust shell
(a sort of quantum mechanical ``matching'' condition)
one would have a correction to the classical no--hair theorem.
\par
The most serious drawback in our approach to Hawking
radiation is the neglect of back--reaction effects.
In fact, one would expect the emitted quanta to alter
the metric outside the dust (since we no longer have vacuum)
and also inside, since the outer shell loses mass.
We hope to return to this and other points.
\par
\bigskip
{\bf Acknowledgements}
\par
\medskip
\noindent
We wish to thank R. Balbinot, R. Bergamini and R. Brout for many
helpful discussions and suggestions.
\appendix
\setcounter{equation}{0}
\section{Useful integrals}
\label{norme}
In the text we defined:
\be
\expec{\hat K^n}\equiv{\bra{\tilde\psi}\,\hat K^{n+2}\,\ket{\tilde\psi}
\over\bra{\tilde\psi}\,\hat K^2\,\ket{\tilde\psi}}
\ ,
\ee
where $\bra{\tilde\psi}\,\hat K^2\,\ket{\tilde\psi}$ 
is the norm of the wave function $K\tilde\psi$.
If one further defines the following differential operator:
\be
D\equiv {1\over2\,\alpha^2}\,{d\over d\,K_{cl}}+K_{cl}
\ ,
\ee
then one has:
\begin{eqnarray}
\bra{\tilde\psi}\,\hat K^n\,\ket{\tilde\psi}
&=&
D^n \int_0^\infty dK\,
e^{-\alpha^2\,(K-K_{cl})^2} 
\nonumber \\
&=&
D^n\left[{\sqrt{\pi}\over 2\,\alpha}\,
\mbox{\rm Erfc}(-\alpha\,K_{cl})\right]
\ .
\end{eqnarray}
In particular one finds: 
\begin{eqnarray}
\bra{\tilde\psi}\,\hat K\,\ket{\tilde\psi}
&=&
{1\over2\,\alpha^2}\,
\left[\sqrt{\pi}\,\alpha\,K_{cl}\,
\mbox{\rm Erfc}(-\alpha\,K_{cl})
+e^{-\alpha^2\,{K_{cl}}^2}\right] \\
& & \nonumber \\
\bra{\tilde\psi}\,\hat K^2\,\ket{\tilde\psi}
&=&{1\over4\,\alpha^3}\,
\left[\sqrt{\pi}\,\left(1+2\,\alpha^2\,{K_{cl}}^2\right)\,
\mbox{\rm Erfc}(-\alpha\,K_{cl})+2\,\alpha\,K_{cl}\,
e^{-\alpha^2\,{K_{cl}}^2}\right] \\
& &\nonumber \\ 
\bra{\tilde\psi}\,\hat K^3\,\ket{\tilde\psi}
&=&
{1\over4\,\alpha^4}\,
\left[\sqrt{\pi}\,\alpha\,K_{cl}\,
\left(3+2\,\alpha^2\,{K_{cl}}^2\right)\,
\mbox{\rm Erfc}(-\alpha\,K_{cl})\right.
\nonumber \\
& &\ \ \ \ \ \ \ \ \left.
+2\,\left(1+\alpha^2\,{K_{cl}}^2\right)\,
e^{-\alpha^2\,{K_{cl}}^2}\right] 
 \\
& & \nonumber \\
\bra{\tilde\psi}\,\hat K^4\,\ket{\tilde\psi}
&=&
{1\over8\,\alpha^5}\,
\left[\sqrt{\pi}\,
\left(3+12\,\alpha^2\,{K_{cl}}^2+4\,\alpha^4\,{K_{cl}}^4
\right)\,\mbox{\rm Erfc}(-\alpha\,K_{cl})\right.
\nonumber \\
& &\ \ \ \ \ \ \ \ \left.
+2\,\alpha\,K_{cl}\,\left(5+2\,\alpha^2\,{K_{cl}}^2\right)\,
e^{-\alpha^2\,{K_{cl}}^2}\right]
\ . 
\end{eqnarray}
\setcounter{equation}{0}
\section{Evaluation of the Fluctuations}
\label{flutt}
From the definition Eq.~(\ref{fase})
one has:
\begin{eqnarray}
\bra{\tilde\chi_{_N}}{\partial^2\over\partial K^2}\ket{\tilde\chi_{_N}}
&=&
\bra{\chi_{_N}}\,\left({\partial\over\partial K}
-\expecl{{\partial\over\partial K}}\right)\,
\left({\partial\over\partial K}
-\expecl{{\partial\over\partial K}}\right)
\,\ket{\chi_{_N}}
\nonumber \\
&=&
\sum\limits_{P\not=N}\,
\bra{\chi_{_N}}\,{\partial\over\partial K}
\,\ket{\chi_{_P}}\,
\bra{\chi_{_P}}\,{\partial\over\partial K}
\,\ket{\chi_{_N}}
\ .
\label{b1}
\end{eqnarray}
From the identity:
\be
{\partial\over\partial K}\left(\hat H_{_M}\,\ket{\chi_{_N}}\right)
={\partial\hat H_{_M}\over\partial K}\,\ket{\chi_{_N}}
+\hat H_{_M}\,{\partial\over\partial K}\ket{\chi_{_N}}
\ ,
\ee
on contracting over $\bra{\chi_{_L}}$ for $L\not= N$
one obtains:
\be
\bra{\chi_{_L}}\,{\partial\over\partial K}
\,\ket{\chi_{_L}}=
{\bra{\chi_{_L}}\,{\partial \hat H_{_M}\over\partial K}
\,\ket{\chi_{_N}}\over
E_{_N}-E_{_L}}
\ ,
\ee
where one has $\hat H_{_M}\,\ket{\chi_{_N}}=E_{_N}\,\ket{\chi_{_N}}=
\hbar\,\mu\,(N+1/2)\,\ket{\chi_{_N}}$.
Further:
\begin{eqnarray}
{\partial \hat H_{_M}\over\partial K}&=&
-{3\over2}\,\left[{{\hat\pi_\phi}^2\over K^4}+\mu^2\,K^3\,\phi^2
\right]
\nonumber \\
&=&{3\over 2}\,{\hbar\,\mu\over K}\,
\left((\hat a^\dagger)^2+{\hat a}^2\right)
\ ,
\end{eqnarray}
where $\hat a^\dagger$ and $\hat a$ are the creation and 
annihilation operators for the scalar field $\hat\phi$: 
\begin{eqnarray}
\hat\phi&=&\sqrt{{\hbar\over2\,\mu\,K^3}}\,
\left(\hat a^\dagger+\hat a\right)
\nonumber \\
& & \\
\hat\pi_\phi&=&i\,\sqrt{{\hbar\,\mu\,K^3\over 2}}\,
\left(\hat a^\dagger-\hat a\right)
\nonumber
\ .
\end{eqnarray}
On substituting into Eq.~(\ref{b1}) one finally has: 
\begin{eqnarray}
\bra{\tilde\chi_{_N}}{\partial^2\over\partial K^2}\ket{\tilde\chi_{_N}}
&=&-\sum\limits_{P\not=N}\,
{1\over(E_{_N}-E_{_P})^2}\,
\bra{\chi_{_N}}\,{\partial \hat H_{_M}\over\partial K}
\,\ket{\chi_{_P}}\,
\bra{\chi_{_P}}\,{\partial \hat H_{_M}\over\partial K}
\,\ket{\chi_{_N}}
\nonumber \\
&=&
-{9\,\hbar^2\,\mu^2\over 4\,K^2}\,
{(N+1)\,(N+2)\over(E_{_N}-E_{_{N+2}})^2}\,
-{9\,\hbar^2\,\mu^2\over 4\,K^2}\,
{N\,(N-1)\over(E_{_N}-E_{_{N-2}})^2}
\nonumber \\
&=&
-{9\over 8\,K^2}\,(N^2+N+1)
\ .
\end{eqnarray}
%
%


\begin{thebibliography}{99}
%
\bibitem{dewitt}
DeWitt B S  1967 {\em Phys. Rev.} {\bf 160} 1113
%
\bibitem{adm}
Arnowitt R Deser S and Misner C 1962 in 
{\em Gravitation: An Introduction to Current Research} 
edited by Witten L. (New York: Wiley) 
%
\bibitem{brout}
Banks T 1985 {Nucl. Phys.} B{\bf 249} 332;
Brout R 1987 {\em Found. Phys.} {\bf 17} 603 and
1987 {\em Z. Phys.} B {\bf 60} 359;
Brout R  and Venturi G  1989 {\em Phys. Rev.} D {\bf 15} 
2436
%
\bibitem{hawking} 
Hawking S W 1974 {\em Nature} (London) {\bf 248} 30;
1975 {\em Comm. Math. Phys.} {\bf 43} 199
%
\bibitem{madsen}
Madsen M S 1988 {\em Class. Quantum Grav.} {\bf 5} 627
%
\bibitem{lund}
Lund F  1973 {\em Phys. Rev.} D {\bf 8} 3253
%
\bibitem{tolman}
Tolman R C  1934 {\em Proc. Nat. Acad. Sci.} {\bf 20}
169
%
\bibitem{oppenheimer}
Oppenheimer J R and Snyder H 1939 {\em Phys. Rev.} {\bf 56}
455
%
\bibitem{bcmn}
Berger B K  Chitre D M  Moncrief V E  and Nutku Y 
1972 {\em Phys. Rev.} D {\bf 5} 2467
%
\bibitem{mtw}
Misner C W  Thorne K S  and Wheeler J A  1973
{\em Gravitation} (San Francesco: Freeman)  
%
\bibitem{israel}
Israel W  1966 {\em Nuovo Cimento} {\bf 44} B 1
%
\bibitem{venturi}
Venturi G  1992 {\em Class. Quantum Grav.} {\bf 9} 1217
%
\bibitem{feynman}
Feynman R P and Hibbs A R 1965 {\em Quantum Mechanics and Path
Integrals} (New York: McGraw--Hill)
%
\bibitem{gibbons}
Gibbons G W and Hawking S W 1977 {\em Phys. Rev. D} {\bf 15}
2752
%
\bibitem{bekenstein}
Bekenstein J D 1974 {\em Lett. Nuovo Cimento} {\bf 11} 476
%
\bibitem{peleg}
Peleg Y 1995 {\em Phys. Lett. B} {\bf 356} 462
%
\bibitem{greensite}
For a general discussion of Ehrenfest's theorem in quantum gravity
see: Greensite J 1991 {\em Nucl. Phys. B} {\bf 351} 749
%
\bibitem{birrell}
Birrell N D and Davies P C W 1982 {\em Quantum Fields in
Curved Space} (Cambridge Univ. Press)
%
\bibitem{unruh}
Unruh W G 1976 {\em Phys. Rev.} D {\bf 14} 870
%
\bibitem{casadio}
Casadio R and Venturi G 1995 {\em Phys. Lett.} A {\bf 199}
33
%
\end{thebibliography}
\end{document}